\documentstyle[12pt]{article}
\newcommand{\nn}{\nonumber}
\def\bbox#1{{\bf #1}}
\def\text#1{{\rm #1}}
\newcommand{\wt}{\widetilde}

\newtheorem{pr}{Proposition}[section]
\def\Journal#1#2#3#4#5#6{{#1}~(#5) #2, {\it #3} {\bf #4}, #6.}
\def\Journaln#1#2#3#4#5#6#7{{#1}~(#5) #2, {\it #3} {\bf #4}(#6), #7.}
\def\Book#1#2#3#4{{#1}~(#4) {\it #2\/}, #3.}
\def\Proc#1#2#3#4#5#6{{#1}~(#6)  #2,~{\it #3\/},~#5,
pp.~#4.}

\def\Epreprint#1#2#3#4{{#1}~(#4) {\it #2\/}, #3.}

\def\JMP{J. Math. Phys.}

\def\LMP{Lett. Math. Phys.}

\def\be{\begin{equation}}
\def\ee{\end{equation}}
\def\bea{\begin{eqnarray}}
\def\eea{\end{eqnarray}}

\renewcommand{\l}{\langle}
\renewcommand{\r}{\rangle}

\begin{document}
\title{Trigonometric Calogero-Moser System
\\as a Symmetry Reduction of KP Hierarchy}
\author{L.V. Bogdanov\thanks
{L.D. Landau ITP, Kosygin str. 2, Moscow 117940, Russia}, ~B.G.
Konopelchenko\thanks {Dipartimento di Fisica dell' Universit\`a and
Sezione INFN, 73100 Lecce, Italy, and IINS, Novosibirsk Branch, Russia}
~and A.Yu. Orlov\thanks {Department of Mathematics, Faculty of Science,
Kyoto University, Kyoto 606-85-02, Japan and Nonlinear wave processes
laboratory,Oceanology Institute, 117218, Krasikova 23, Moscow, Russia.}}
\date{}
\maketitle
\begin{abstract}
Trigonometric non-isospectral flows are defined for KP hierarchy. It is
demonstrated that symmetry constraints of KP hierarchy associated with
these flows give rise to trigonometric Calogero-Moser system.
\end{abstract}
\section{Introduction}
This paper may be considered as a sequel of the work \cite{Chg99}, where
it was shown that rational Calogero-Moser system can be obtained by a
symmetry constraint of KP hierarchy. Here we show that a simple
generalization of the scheme leads to {\em trigonometric} Calogero-Moser
system. We describe the corresponding symmetries and symmetry constrains
in the framework of analytic-bilinear approach to integrable hierarchies
\cite{AB1,AB2,mybook} (the primary objects in this approach are
Cauchy-Baker-Akhiezer (CBA) function and Hirota bilinear identity for
it), as well as in terms of free fermionic fields \cite{Date}.

\section{Hirota Identity and KP Hierarchy}
First we give a sketch of the picture of generalized KP hierarchy in
frame of analytic-bilinear approach; for details we refer to
\cite{AB1,AB2,mybook}.

The formal starting point is Hirota bilinear identity
for Cauchy-Baker-Akhiezer function,
\bea
\oint \chi(\nu,\mu;g_1)g_1(\nu)g_2^{-1}(\nu)
\chi(\lambda,\nu;g_2)d\nu=0\,\quad \lambda\,,\mu\in D.
\label{HIROTA0}
\eea
Here $\chi(\lambda,\mu;g)$ (the Cauchy kernel)
is a function of two complex variables
$\lambda ,\mu\in \bar D$, where $D$ is a unit disc,
and a functional of the loop group
element $g\in\Gamma^+$, i.e., of a
complex-valued function analytic and having
no zeros in $\bbox{C}\setminus D$, equal to 1 at infinity;
the integration goes over the
unit circle . By definition, the function $\chi(\lambda,\mu)$
possesses the following analytical properties: as $\lambda
\rightarrow\mu$, $\chi\rightarrow (\lambda-\mu)^{-1}$ and
$\chi(\lambda,\mu)$ is an analytic function of two variables
$\lambda,\mu \in \bar D$ for $\lambda\neq\mu$. The function
$\chi(\lambda,\mu;g)$ is a solution to (\ref{HIROTA0}) if it possesses
specified analytic properties and satisfies (\ref{HIROTA0}) for all
$\lambda,\mu \in D$ and some class of loops $g\in\Gamma^+$.

Parametrization of $\Gamma^+$ in terms of standard KP variables
$$
g(\lambda)=g(\lambda,\bbox{x})=\exp\left(\sum_{n=1}^\infty
\lambda^{-n}x_n\right)
$$
gives an opportunity to consider functionals of $g\in\Gamma^+$ as
functions of an infinite set of KP variables $\bbox{x}$.

In  another form, more similar to standard Hirota bilinear identity,
the identity (\ref{HIROTA0}) can be written as
\be
\oint \psi(\lambda,\nu;g_2)
\psi(\nu,\mu;g_1)d\lambda=0 ,
\ee
where
$$
\psi(\lambda,\mu,g)=g(\lambda)\chi(\lambda,\mu,g)g^{-1}(\mu).
$$
We call the function $\psi(\lambda,\mu;g)$ a Cauchy-Baker-Akhiezer
function.

Hirota bilinear identity (\ref{HIROTA0})
incorporates the standard Hirota bilinear identity
for the Baker-Akhiezer (BA) and dual (adjoint) Baker-Akhiezer function
of the KP hierarchy.
Indeed, let us introduce
these functions by the formulae
\bea
\psi(\lambda;g)=g(\lambda)\chi(\lambda;0),
\nn\\
\wt\psi(\mu;g)=g^{-1}(\mu)\chi(0;\mu).
\nn
\eea
Then for Baker-Akhiezer function $\psi(\lambda;g)$ and dual
Baker-Akhiezer function $\wt\psi(\mu;g)$, taking the identity
(\ref{HIROTA0}) at $\lambda=\mu=0$, we get the usual form of Hirota
bilinear identity
\be
\oint \wt\psi(\nu;g_2)
\psi(\nu;g_1)d\nu=0.
\ee
The only difference with the standard setting here is that we define BA
and dual BA function in the neighborhood of zero, not in the
neighborhood of infinity.

There are three different types of integrable equations implied by
identity (\ref{HIROTA0}), that correspond to the KP hierarchy in the
usual form (in terms of potentials), to the modified KP hierarchy and to
the hierarchy of the singular manifold equations. They arise for
different types of functions connected with the Cauchy-Baker-Akhiezer
function satisfying Hirota bilinear identity (see the derivation in
\cite{AB1}, \cite{AB2}, \cite{mybook}).\\
1.On the first level, we have hierarchy of equations for the diagonal of
the regularized Cauchy kernel taken at zero (the potential)
\bea
&& u(g)=\chi_r(0,0;g),
\nn\\
&& \chi_r(\lambda,\mu;g)=
\chi(\lambda,\mu;g)- (\lambda-\mu)^{-1}\,.
\nn
\eea
The first equation of this hierarchy is a potential form of  KP equation
\begin{equation}
\partial_x\left(u_t-\mbox{${1\over 4}$}u_{xxx}+
\mbox{${3\over2}$}(u_x)^2\right)=\mbox{${3\over4}$}
u_{yy}\,,
\label{KP00}
\end{equation}
where $x=x_1$, $y=x_2$, $t=x_3$, which reduces to standard KP equation
for the function $v=-2\partial_x u$.
\\2.On the second level, there are the equations for the
Baker-Akhiezer and dual Baker-Akhiezer type wave functions (the modified
equations)
$$
\Psi(g)=\int\psi(\lambda,g) \rho(\lambda)d\lambda\,,
$$
$$
\widetilde \Psi(g)=
\int\widetilde\rho(\mu)\widetilde\psi(\mu,g) d\mu\,.
$$
\\
3.On the third level we have the equations for the Cauchy-Baker-Akhiezer
type wave function
\bea
\Phi(g)=\oint\!\!\oint (\psi_r(\lambda,\mu;g))
\rho(\lambda)
\wt\rho(\mu) d\lambda\,d\mu,
\nn
\eea
where $\rho(\lambda)$, $\wt\rho(\mu)$ are some arbitrary weight
functions.

\section{Nonisospectral Symmetries}
The (isospectral) dynamics defined by Hirota bilinear identity
(\ref{HIROTA0}) is connected with operator of multiplication by loop
group element $g\in
\Gamma^+$; this dynamics can be interpreted in terms of commuting flows
corresponding to infinite number of `times' $x_n$. A general idea of
introduction additional (in general, non-commutative) symmetries is to
consider more general operators $\hat R$ on the unit circle. Let us
introduce symmetric bilinear form
$$
(f|g)=\oint f(\nu)g(\nu)d(\nu).
$$
In terms of this form identity (\ref{HIROTA0})
looks like
\bea
(\chi(\dots,\mu;g_1)g_1(\dots)|
g_2^{-1}(\dots)\chi(\lambda,\dots;g_2))=0\,\quad \lambda\,,\mu\in D,
\label{HIROTA01}
\eea
or, for Cauchy-Baker-Akhiezer function $\psi(\lambda,\mu;g)$,
\bea
(\psi(\dots,\mu;g_1)|
\psi(\lambda,\dots;g_2))=0\,\quad \lambda\,,\mu\in D,
\label{HIROTA02}
\eea
where by dots we denote the argument which is involved into
integration.
Let some CBA function $\psi(\lambda,\mu;g)$ satisfying Hirota
bilinear identity be given.
We define symmetry transformation connected with
arbitrary invertible
linear operator $\hat R$ in the space of functions on the
unit circle by the equations
\bea
(\wt\psi(\dots,\mu;g_1)|\hat R|
\psi(\lambda,\dots;g_2))=0,
\nn\\
(\psi(\dots,\mu;g_1)|\hat R^{-1}| \wt\psi(\lambda,\dots;g_2))=0. \nn
\eea
It is possible to show that if both these equations for the transformed
CBA function $\wt\psi(\lambda,\mu;g)$ are solvable, then the solution
for them is the same (and unique), and it satisfies identity
(\ref{HIROTA02}). In this case the symmetry transformation connected with
operator $\hat R$ is correctly defined. It is also possible to define
one-parametric groups of transformations by the equation
\bea
(\psi(\dots,\mu;g_1,\Theta_1)|\exp((\Theta_1-\Theta_2)\hat r)|
\psi(\lambda,\dots;g_2,\Theta_2))=0, \label{H1}
\eea
where $\Theta$ is a parameter (non-isospectral `time'). Taking the
generators $\hat r_{mn}=\lambda^{-m}\partial_\lambda^n$, we get
noncommutative symmetries in the form proposed by Orlov and Shulman
\cite{Orlov}.

In the work \cite{Chg99} non-isospectral symmetries connected with
operators with degenerate kernel were considered (similar symmetries
were studied in \cite{Orlov2}). In particular, generators with the kernel
of the form
\be
r_{\alpha\beta}(\nu,\nu')={2\pi\text{i}} \delta(\alpha-\nu)
\delta(\beta-\nu'),
\ee
where $\alpha,\beta$ belong to the unit circle, were used.

More general case of generators
\be
r_{\rho\wt\rho}(\nu,\nu')={2\pi\text{i}}\wt\rho(\nu')\rho(\nu),
\label{rho}
\ee
was also studied, but for simplicity it was put
$$
(\wt\rho|\rho)=0.
$$
The crucial point for this work is to {\em generalize this condition},
and consider degenerate operators with nonzero pairing of factors
$$
{2\pi\text{i}}(\wt\rho|\rho)=h,
$$
or, more generally,
\bea
r_{h}(\nu,\nu')= {2\pi\text{i}}
\sum_{i=1}^N\wt\rho_i(\nu')\rho_i(\nu),\nn\\
{2\pi\text{i}}(\wt\rho_i|\rho_j)=h\delta_{ij},
\label{trigogen}
\eea
where $h$ is some constant. We will show that generators of these form
define trigonometric nonisospectral flows.

\subsection{Trigonometric Flows}
Using simple identity
$$
\hat r_h^2=h \hat r_h,
$$
we get the formula
\be
\exp(\Theta \hat r_h)=1+\hat r_h\frac{e^{\Theta h}-1}{h}.
\ee
Then, performing
integration in the equation (\ref{H1}) taken for $g_1=g_2$, which in
this case reads
\bea
\oint\oint d\nu d\nu'
\psi(\nu,\mu;g,\Theta_1)(\delta(\nu-\nu')+
\frac{e^{(\Theta_1-\Theta_2) h}-1}{h}
{2\pi\text{i}}
\sum_{i=1}^N\wt\rho_i(\nu')\rho_i(\nu))
&&
\nn\\
\times \psi(\lambda,\nu';g,\Theta_2))=0,&& \label{H2} \eea we get
equation for the CBA function
\bea &&
\psi(\lambda,\mu;\bbox{x},\Theta+\Delta\Theta) =
\psi(\lambda,\mu;\bbox{x},\Theta)
\nn\\
&&
\qquad
+\frac{e^{h\Delta\Theta }-1}{h}\sum_{i=1}^N
\wt\phi_i(\lambda;\bbox{x},\Theta)
\phi_i(\mu;\bbox{x},\Theta+\Delta
\Theta),
\label{Delta}
\eea
or, in differential form,
\bea
\partial_\theta\psi(\lambda,\mu;\bbox{x},\Theta)=
\sum_{i=1}^N
\wt\phi_i(\lambda;\bbox{x},\Theta)
\phi_i(\mu;\bbox{x},\Theta),
\label{Delta11}
\eea
where
\bea
\wt\phi_i(\lambda;\bbox{x},\Theta)=
\oint d\mu \psi(\lambda,\mu;\bbox{x},\Theta)\wt\rho(\mu),
\nn\\
\phi_i(\mu;\bbox{x},\Theta)=
\oint d\lambda \psi(\lambda,\mu;\bbox{x},\Theta)\rho(\lambda).
\nn
\eea

It is possible to resolve equation (\ref{Delta}) and express
$\psi(\lambda,\mu;\bbox{x},
\Theta_{\alpha\beta}+\Delta\Theta_{\alpha\beta})$ through
$\psi(\lambda,\mu;\bbox{x},\Theta_{\alpha\beta})$. First we integrate
equation (\ref{Delta}) with the weight function $\rho_k(\lambda)$ and get
\bea &&
\phi_k(\mu;\bbox{x},\Theta+\Delta\Theta) =
\phi_k(\mu;\bbox{x},\Theta)
\nn\\
&&
\qquad
+\frac{e^{h\Delta\Theta }-1}{h}\sum_{i=1}^N
\Phi_{ki}(\bbox{x},\Theta)
\phi_i(\mu;\bbox{x},\Theta+\Delta
\Theta),
\label{Delta01}
\eea
where
\bea
&&
\Phi_{ij}(g)=\oint\!\!\oint (\psi(\lambda,\mu;g)-(\lambda-\mu)^{-1})
\rho_i(\lambda)
\wt\rho_j(\mu) d\lambda\,d\mu
\nn\\
&&\qquad\qquad - 2\pi\text{i} (\rho_i^\text{out}|\wt\rho_j^\text{in}),
\nn
\eea
and
\bea
&&
f^\text{in}(\lambda)={1\over 2\pi\text{i}}\oint d\mu f(\mu)
(\mu-\lambda^\text{in})^{-1}
\nn\\&&\qquad\qquad
={1\over 2\pi\text{i}}\text{v.p.}
\oint d\mu f(\mu)
(\mu-\lambda)^{-1}+{1\over 2}f(\lambda),
\nn\\&&
f^\text{out}(\lambda)=-{1\over 2\pi\text{i}}\oint d\mu f(\mu)
(\mu-\lambda^\text{out})^{-1}
\nn\\&&\qquad\qquad
=-{1\over 2\pi\text{i}}\text{v.p.}
\oint d\mu f(\mu)
(\mu-\lambda)^{-1}+{1\over 2}f(\lambda).
\nn
\eea
In matrix form,
\bea
&& |\phi(\mu;\bbox{x},\Theta+\Delta\Theta)\rangle
\nn\\&&\qquad\qquad=
|\phi(\mu;\bbox{x},\Theta)\rangle+
\frac{e^{h\Delta\Theta
}-1}{h}
\Phi(\bbox{x},\Theta)|\phi(\mu;\bbox{x},\Theta+\Delta\Theta)\rangle.
\nn
\eea
Resolving this equation with respect to
$|\phi(\Theta+\Delta\Theta)\rangle$, we obtain
\bea
|\phi(\mu;\bbox{x},\Theta+\Delta\Theta)\rangle=(I-
\frac{e^{h\Delta\Theta }-1}{h}\Phi(\bbox{x},\Theta))^{-1}|
\phi(\mu;\bbox{x},\Theta)\rangle.
\label{Delta1}
\eea
Substituting (\ref{Delta1}) into (\ref{Delta}), we finally get
\bea &&
\psi(\lambda,\mu;\bbox{x},\Theta+\Delta\Theta)=
\psi(\lambda,\mu;\bbox{x},\Theta)
\nn\\&&
\quad
+
\frac{e^{h\Delta\Theta }-1}{h}\langle\wt\phi(\lambda;\bbox{x},\Theta)|
(I-\frac{e^{h\Delta\Theta }-1}{h}\Phi)^{-1}|
\phi(\mu;\bbox{x},\Theta)\rangle.
\label{Delta2}
\eea
The formula (\ref{Delta2}) explicitly defines {\em discrete
nonisospectral symmetry} of KP hierarchy in terms of CBA function.

In particular, this formula expresses the function
$\psi(\lambda,\mu;\bbox{x},\Theta)$ through the initial data
$\psi_0(\lambda,\mu;\bbox{x})=
\psi(\lambda,\mu;\bbox{x},\Theta=0)$, thus giving explicit
formula for the action of non-isospectral flow connected with the
generator (\ref{trigogen}) on the CBA function. This flow appears to be
{\em trigonometric}, because the CBA function and other objects of the
hierarchy defined through it (potential, wave functions) depend
rationally on $\exp(h\Theta)$.

Using the formula (\ref{Delta2}), it is also possible to get the action
of the trigonometric flow on the $\tau$-function. Using simple identity
$$
\det (I+|f\rangle\langle g|)=\langle g|f\rangle
$$
and determinant formula for the transformation of CBA function under the
action of a rational loop (see \cite{mybook}),
\bea
\psi_0(\alpha,\beta;\bbox{x}+[\mu]-[\lambda])
={\det
\left(
\begin{array}{cc}
\psi_0(\lambda,\mu;\bbox{x})&\psi_0(\lambda,\beta;\bbox{x})\\
\psi_0(\alpha,\mu;\bbox{x})&\psi_0(\alpha,\beta;\bbox{x})
\end{array}
\right)
\over \psi_0(\lambda,\mu;\bbox{x})},
\eea
we get another representation of the formula (\ref{Delta2}),
\bea
\psi(\lambda,\mu;\bbox{x},\Theta)=
\psi_0(\lambda,\mu;\bbox{x})
{\det(I-\frac{e^{h\Theta }-1}{h}\Phi_0(\bbox{x}+[\mu]-[\lambda]))
\over \det(I-\frac{e^{h\Theta }-1}{h}\Phi_0(\bbox{x}))}.
\eea
Comparing this formula with the formula connecting the CBA function and
the $\tau$-function (which in fact defines the $\tau$-function through
the CBA function)
\bea
\psi(\lambda,\mu,\bbox{x})=
{g(\lambda,\bbox{x})g(\mu,\bbox{x})^{-1}}{1\over
\lambda-\mu}{\tau(\bbox{x}+[\mu]-[\lambda]) \over\tau(\bbox{x})},
\eea
we come to the conclusion that the $\tau$-function corresponding to the
transformed CBA function $\psi(\lambda,\mu;\bbox{x},\Theta)$ is given by
the expression
\bea
\tau(\bbox{x},\Theta)=
\tau_0(\bbox{x})\det\left(I-\frac{e^{h\Theta }-1}{h}\Phi_0(\bbox{x})
\right).
\label{tautrans}
\eea
Thus we have explicitly defined action of non-isospectral symmetry with
the generator (\ref{trigogen}) on KP $\tau$-function. This formula also
defines the evolution of KP potential $u(\bbox{x})$,
\bea &&
u(\bbox{x},\Theta)=\psi_r(0,0;\bbox{x},\Theta)=
-\partial_x\ln\tau(\bbox{x},\Theta)
\nn\\&&\qquad\qquad
= u_0(\bbox{x})-
\partial_x\ln \det\left(I-\frac{e^{h\Theta }-1}{h}\Phi_0(\bbox{x})
\right),
\label{transu}
\eea
where $x=x_1$ (it is easy to get this formula directly from
(\ref{Delta2})).

\subsection{M\"obius-type Symmetry}
The transformation of the matrix $\Phi$ under the action of
trigonometric nonisospectral flow is especially simple. According to
(\ref{Delta2}), it looks like
\bea
&&
\Phi(\bbox{x},\Theta)= {\Phi_0(\bbox{x})e^{h\Theta}
\over 1-\frac{e^{h\Theta }-1}{h}\Phi_0(\bbox{x})},
\label{moebius}
\eea
and it is nothing more than matrix M\"obius-type transformation.
Differential equation defining this transformation is
\bea
{\partial\Phi(\bbox{x},\Theta)\over\partial\Theta}=
\Phi^2(\bbox{x},\Theta)+h \Phi(\bbox{x},\Theta)
\label{moebiusgen}
\eea
The difference with the work \cite{Chg99} is that we consider generic
M\"obius-type one-parametric flow, which is trigonometric.

We would like to recall (see \cite{Chg99})  that transformation of the KP
potential $u$ corresponding to matrix inversion $\Phi^{-1}$ looks like
\be
u(\Phi^{-1})=u(\Phi)-\partial_x\ln\det \Phi,
\label{transmulti2}
\ee
and it represents a composition formula for several binary B\"acklund
transformations (this formula can also be derived from (\ref{transu}),
(\ref{moebius}) in the limit $\Theta\rightarrow\infty$). Taking into
account that the transformations $\Phi+C$, $A\Phi B$, where $A,B,C$ are
constant matrices, correspond to identical transformation of the
potential, the formula (\ref{transmulti2}) is sufficient to define the
transformation of potential corresponding to arbitrary matrix M\"obius
transformation of $\Phi$. Thus, to the derive formula (\ref{transu}), it
is enough to fix the generator of one-parametric subgroup of the M\"obius
group (\ref{moebiusgen}) and to use the formula  (\ref{transmulti2}).

\section{Symmetry Constraints and Calogero-Moser System}
Now, when we have identified trigonometric nonisospectral flows, it is
quite straightforward to interpret Calogero-Moser system as a symmetry
constraint of KP hierarchy.
\begin{pr}
Let us impose the following symmetry constraint:
\be
\partial_\Theta u(\bbox{x},\Theta)=
\partial_x u(\bbox{x},\Theta).
\label{constraint}
\ee
Then the dependence $u(\bbox{x},\Theta)$ on $x$ is trigonometric, and the
motion of poles of $u(\bbox{x},\Theta)$ in the $x$-plane with respect to
the `time' $y=x_2$ is described by trigonometric Calogero-Moser system.
\label{uconstraint}
\end{pr}
{\bf Proof.} The dependence of KP potential on $\Theta$ is explicitly
given by the formula (\ref{transu}) and it is trigonometric (rational in
$\exp(h\Theta)$). Thus a constraint enforces trigonometric dependence of
$u(\bbox{x},\Theta)$ on $x$, that, according to \cite{Krichever}, leads
to trigonometric Calogero-Moser system
\be
\partial_y^2 x^i=4\sum_{j\neq i}V'(x^i-x^j),
\quad V(x)= {1\over\sinh^2x},\quad
V'(x)=\partial_x V(x).
\ee
Solutions to this system, due to the formula (\ref{transu}), are defined
through the eigenvalues of the matrix $\Phi(\bbox{x})$.\\

We would like to make a remark clarifying the relation between constraint
(\ref{constraint}) and standard constrained KP hierarchy (see
\cite{KS0}, \cite{KS}),
which is defined by the condition
\be
\partial_x u(\bbox{x})=\sum_{i=1}^N\Psi_i(\bbox{x})\wt\Psi_i(\bbox{x}),
\label{constraint1}
\ee
and represents multicomponent AKNS-type system for the wave functions.
The KP potential $u(\bbox{x},\Theta)$ defined by the formula
(\ref{transu}) satisfies differential equation
\be
\partial_\Theta u(\bbox{x},\Theta)=\partial_x
\text{tr}|\Phi(\bbox{x},\Theta)|=
\sum_{i=1}^N \Psi_i(\bbox{x},\Theta)\wt\Psi_i(\bbox{x},\Theta)
\ee
(it is easy to check it directly or derive from (\ref{Delta11})). Thus,
the standard constraint is given by the condition (\ref{constraint})
taken at the single point $\Theta=0$ (or at some fixed point). Therefore
the constraint (\ref{constraint}) is a stronger constraint, and
potentials $u(\bbox{x},\Theta)$ satisfying this constraint also satisfy
condition (\ref{constraint1}) (for all $\Theta$).

\section{Fermionic Fields and Calogero-Moser System }
The results we have presented can be reformulated using the KP theory in
terms of fermionic fields \cite{Date}. Now let us review some facts from
this theory. We have fermionic fields
$$\psi(z)=\sum_{k\in Z}
\psi_k z^k,\quad\psi^*(z)=\sum_{k\in Z} \psi^*_k z^{-k-1},$$
where fermionic
operators satisfy the canonical anti-commutation relations: \be
\label{antikom} [\psi_m,\psi_n]_+=[\psi^*_m,\psi^*_n]_+=0;\qquad
[\psi_m,\psi^*_n]_+=
\delta_{mn} . \ee Let us introduce left and right vacuums by the
properties:
\begin{eqnarray}\label{vak}
\psi_m |0\r=0 \qquad (m<0),\qquad \psi_m^*|0\r =0 \qquad (m \ge 0), \nn\\
\l 0|\psi_m=0 \qquad (m\ge 0),\qquad \l 0|\psi_m^*=0 \qquad (m<0).\nn
\end{eqnarray}
Note that the subscript $*$ does not denote the complex conjugation. The
vacuum expectation value is defined by relations:
\begin{eqnarray}
\l 0|1|0 \r=1,\quad \l 0|\psi_m\psi_m^* |0\r=1\quad m<0, \quad  \l
0|\psi_m^*\psi_m |0\r=1\quad m\ge 0 ,
\nn
\end{eqnarray}
\bea
\l 0|\psi_m\psi_n |0\r=
\l 0|\psi^*_m\psi^*_n |0\r=0,\quad \l 0|\psi_m\psi_n^* |0\r=0 \quad m\ne n.
\nn
\eea

Let us denote $\widehat{gl}(\infty)= {\itshape{Lin}}\{
1,:\psi_i\psi_j^*:|i,j \in Z\}$, with usual normal ordering
$:\psi_i\psi_j^*:=\psi_i\psi_j^*-\l 0|\psi_i\psi_j^*|0\r$. We define the
operator $g$ which is an element of the group $\widehat{GL}(\infty)$
corresponding to the infinite dimensional Lie algebra
$\widehat{gl}(\infty)$. The $\tau$-function of the KP equation is
sometimes defined as
\be\label{taucorKP}
\tau_{KP}(M,{\bf t})=
\langle M|e^{H({\bf
t})}g|M \rangle ,\quad M \in Z,
\ee
where ${\bf t}=(t_1,t_2,\dots)$
 is the set of higher KP times. $H({\bf t}) \in {\widehat {gl}}(\infty)$
is given by
\bea
H({\bf t})=\sum_{n=1}^{+\infty} t_n H_n ,
\quad H_n=\frac{1}{2\pi
i}\oint :z^{n}\psi(z)\psi^*(z): dz.
\nn
\eea
According to
\cite{Date},\cite{UT} the integer $M$ in (\ref{taucorKP}) plays the role of discrete
Toda lattice variable and defines the following charged vacuums
\begin{eqnarray}
\l M|=\l 0|\Psi^{*}_{M},\qquad |M\r=\Psi_{M}|0\r ,
\nn
\end{eqnarray}
\begin{eqnarray}
\Psi_{M}=
\psi_{M-1}\cdots\psi_1\psi_0 \quad M>0,\qquad \Psi_{M}=
\psi^{*}_{M}\cdots\psi^{*}_{-2}\psi^{*}_{-1}\quad M<0 , \nonumber\\
\Psi^{*}_{M}=\psi^{*}_{0}\psi^{*}_1\cdots\psi^{*}_{M-1} \quad M>0,\qquad
\Psi^{*}_{M}=\psi_{-1}\psi_{-2}\cdots\psi_{M}\quad M<0 .
\nn
\end{eqnarray}

The Baker-Akhiezer function and the conjugated one are
\be\label{taucor1}
w(M,{\bf t};k)=\frac {\l M+1|e^{H({\bf t})}
\psi(k)g|M \r } {\l M |e^{H({\bf t})}g|M \r }, \ee
\be\label{w*} w^*(M,{\bf t};k)=\frac {\l M-1|e^{H({\bf
t})}\psi^*(k)g|M \r } {\l M |e^{H({\bf
t})}g|M \r }.
\ee

\subsection{Trigonometric Flows}

Let us consider the linear combinations of fermionic fields:
\be\label{AA*}
A_i=\oint c_i(k)\psi(k)dk,\quad A^*_i=\oint
c^*_i(k)\psi^*(k)dk, \quad i=1,2,\dots , N,
\ee
which satisfy the relations
\be\label{h}
A_iA^*_j+A^*_jA_i=h\delta_{ij},
\ee
where $h \in C$ is a finite number, $c_i(k),c^*_i(k) \in C$.

Then we introduce a $\tau$-function which depends on a time $\beta$ in
the following way:
\be\label{tauflow}
\tau(M,{\bf t},\beta )= \l
M|e^{H({\bf t})}\prod_{i=1}^Ne^{\beta A_i A^*_i} g|M\r
.
\ee
This dependence describes a special one-parametric Backlund
transformation of the $\tau$-function $\tau(M,{\bf t},\beta=0)$. This is
the fermionic version of N-fold ``Zakharov-Shabat dressing depending on
functional parameters'' \cite{ZM} of a given (nonvacuum) solution, see
\cite{Orlov2}.
\begin{pr}
$\tau(M,{\bf t},\beta )$ is a trigonometric function of $\beta$.
\end{pr}
{\bf Proof.} For each pair $A=A_i,\;A^*=A_i^*$ we have
\be e^{\beta A
A^*}=1+\beta AA^* +\frac {\beta^2}{2!} AA^*AA^*+...= 1+ AA^*
\frac{e^{\beta h}-1}{h}.
\ee
Thus $\tau$-function is a trigonometric function of $\beta$.

\subsection{Trigonometric Calogero-Moser System (TCMS)}

\begin{pr}
Let us impose the following symmetry constraint ($x=t_1$):
\be\label{x=beta} \frac {\partial \tau}{\partial \beta}=\frac {\partial
\tau}{\partial x}. \ee Then the motion of zeroes of
$\tau(M,{\bf t},\beta )$
in the $x$-plane is described by trigonometric
Calogero-Moser system.
\label{tauconstraint}
\end{pr}
{\bf Proof.} It follows from (\ref{x=beta}) that $\tau$-function
(\ref{tauflow}) is a trigonometric function of the variable $x$. In turn
it is known
\cite{Krichever}, \cite{Shiota} that zeroes of $\tau$-function which is the
rational
expression of $e^{xh}$, $h$ is a constant, are described by RCMS.
\\
\\
Now let
\be\label{AA*1} a_i({\bf t})=\oint c_i(k)w({\bf t};k)dk,
\quad a^*_i({\bf t})=\oint
c^*_i(k)w^*({\bf t};k)dk, \quad i=1,2,\dots , N.
\ee
\begin{pr}
Due to the formulae (\ref{AA*1}), (\ref{taucor1}), (\ref{w*}) the
$\tau$-function (\ref{tauflow}), (\ref{x=beta}) is also the
$\tau$-function of the following multicomponent Schr\"o\-din\-ger
equation
\be
(\partial_{t_2}-\partial_{t_1}^2-\sum_i^N a_ia^*_i)a_j=0,\quad
(\partial_{t_2}+\partial_{t_1}^2+\sum_i^N a_ia^*_i)a^*_j=0.
\ee
\end{pr}
See \cite{Orlov2} for proof.

It follows from the results of
\cite{Grinevich},\cite{Orlov2} that
$\beta$ can be identified with $\Theta$ of {\em Sections 3,4}. Compare
also Proposition \ref{uconstraint} and Proposition \ref{tauconstraint}

The open problem is the problem of completeness of TCMS
solutions which we get via the symmetry reduction among all TCMS
solutions.

\section*{Acknowledgments}
The work of LB was supported in part by the Russian Foundation for Basic
Research grants No 98-01-00525 and 00-15-96007 and by INTAS grant
99-1782. AO thanks T. Shiota and A. Mironov for the discussion.

\end{document}